\documentclass{article} 
\usepackage{spconf,amsmath,graphicx}
\usepackage{booktabs}
\usepackage{makecell}
\usepackage{dblfloatfix}
\usepackage{xcolor}
\usepackage{multicol}

\usepackage[]{algorithm2e}

\usepackage{blindtext}
\usepackage{tikz}
\usetikzlibrary{calc}

\DeclareMathOperator*{\argmin}{arg\,min}

\title{Unsupervised Discriminative Learning of Sounds for Audio Event Classification}
\name{Sascha Hornauer$^1$, Ke Li$^1$, Stella X. Yu$^1$, Shabnam Ghaffarzadegan$^2$, Liu Ren$^2$}
\address{UC Berkeley$^1$, Bosch Research North America$^2$}
\begin{document}
\begin{figure*}
\copyright 2021 IEEE. Personal use of this material is permitted. Permission from IEEE must be obtained for all other uses, in any current or future media, including reprinting/republishing this material for advertising or promotional purposes, creating new collective works, for resale or redistribution to servers or lists, or reuse of any copyrighted component of this work in other works.

\end{figure*}

\maketitle
\tikz[remember picture, overlay]
    \node [rotate=90] at ($(current page.west)+(1,0)$)
      {\footnotesize ICASSP 2021 - 2021 IEEE International Conference on Acoustics, Speech and Signal Processing (ICASSP) | 978-1-7281-7605-5/20/\$31.00 ©2021 IEEE | DOI: 10.1109/ICASSP39728.2021.9413482};
      
\begin{abstract}
    Recent progress in network-based audio event classification has shown the benefit of pre-training models on visual data such as ImageNet. While this process allows knowledge transfer across different domains, training a model on large-scale visual datasets is time consuming. On several audio event classification benchmarks, we show a fast and effective alternative that pre-trains the model unsupervised, only on audio data and yet delivers on-par performance with ImageNet pre-training. Furthermore, we show that our discriminative audio learning can be used to transfer knowledge across audio datasets and optionally include ImageNet pre-training.
\end{abstract}

\vspace{-0.3cm} 
\section{Introduction}

Deep learning for audio event detection and classification benefits from large datasets. 
Despite unlimited access to audio data (e.g. YouTube, Freesound, etc), labeling audio events is labor intensive and noisy due to ambiguity in start and end times and short duration of some audio events.

On the other hand, due to similarities of the most commonly used audio features (i.e. spectrograms) to images, it is possible to benefit from advances in the image and video domain.
Recent work shows improved performance when pre-training models on pretext tasks such as image classification or video based prediction \cite{cramer2019look, yang2020telling}. 

However, using image/video pre-training is very time demanding due to the size of visual data.
Also, for every architectural change, this time-demanding pre-training needs to be repeated.
Finally, it limits network design since the feature extractor has to be able to process image data which might not be necessarily suitable for an audio task.

Furthermore, for audio applications on embedded devices, such as voice assistants, it is desirable to be able to improve the model performance on the edge over time through fine-tuning on new recorded data. The need of on the edge computation might be due to privacy concerns, to avoid sending users' audio data to the Internet, or due to missing network availability. As a result, task performance needs to improve fast with few epochs and little data on the device. Large network models, as used for image data, may be too computationally expensive for many devices.
\begin{figure}[ht]
    \centering
    \includegraphics[width=1.0\columnwidth]{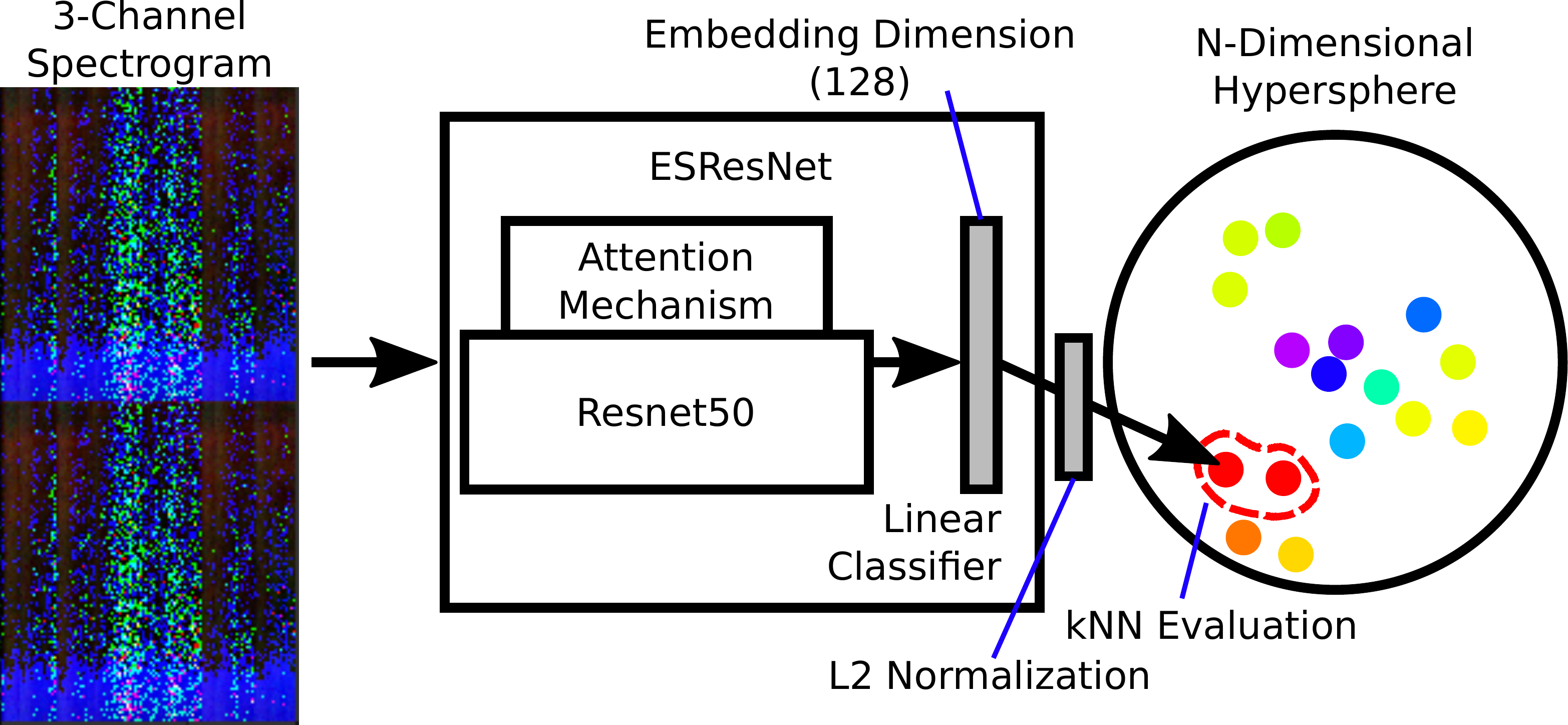}
    \caption{Embedding of spectrograms into the feature space with ESResNet. Stereo channels are stacked for illustration. First the network learns to embed similar spectrograms close together on a hypersphere. Second, sound event classification is trained, starting with the pre-trained network.}
    \label{fig:schema_training}
\end{figure}

We present a pre-training method that accelerates network fine-tuning on the task of sound classification while being itself fast, efficient and versatile. Compared with the state-of-the-art approach ESResNet \cite{guzhov2020esresnet}, we show how our method achieves competitive results and significantly outperforms training from scratch. On one benchmark we even outperform state-of-the-art pre-training in early epochs. 

We are faster at pre-training because we need only three audio datasets which combined have a fraction of the size of ImageNet. We focus on achieving fast results in very few epochs for edge computation and do not aim to go beyond state of the art performance on sound classification after extensive training. For a fair comparison we apply our method to the ESResNet codebase. 

\begin{figure*}[ht]
    \centering
    \includegraphics[width=1.0\textwidth]{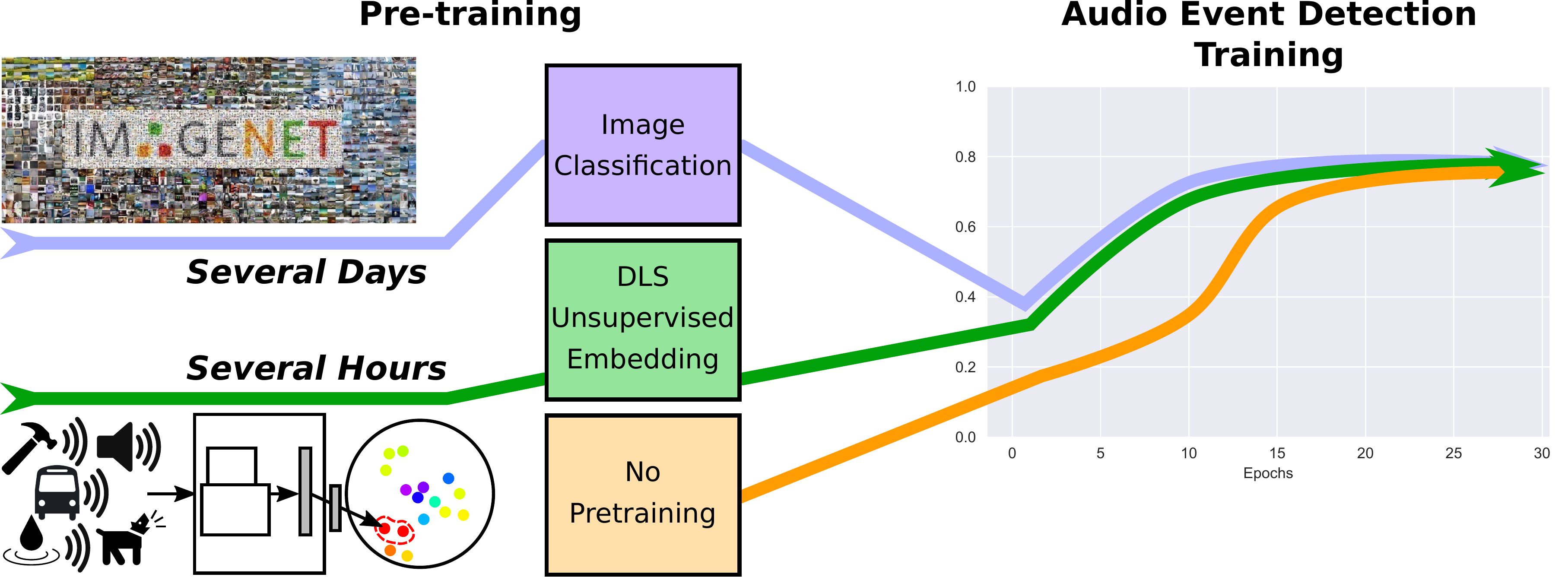}
    \caption{Discriminative Learning of Sounds (DLS) for Audio Event Classification. We compare the same network, pre-trained either on our proxy-task, ImageNet image classification or not at all. ImageNet pre-training takes several days on common GPUs. DLS can train unsupervised, only on sound data and within a few hours. When fine-tuning on Audio Event Detection, DLS stays on par with ImageNet pre-training over many epochs. It constitutes an efficient alternative, especially useful for edge computing or to accelerate the design phase of novel network architectures.}
    \label{fig:overview}
\end{figure*}
With unsupervised training on a pretext task, using only audio data, we also avoid the need for labels. By using Non-Parametric Instance-level Discrimination (NPID) \cite{wu2018unsupervised} to train ESResNet on audio datasets we learn features beneficial for downstream audio classification tasks, illustrated in fig. \ref{fig:schema_training}. This allows us to integrate all data seamlessly across datasets and after deployment to train on novel unlabeled audio data. We call this approach Unsupervised Discriminative Learning of Sounds (DLS) and give an overview in fig. \ref{fig:overview}.

\section{Related Work}

\textbf{Audio Event Detection.} Audio event detection was largely improved in the last decade by leveraging new available datasets \cite{7100934}. We use: ESC-50, its subset ESC-10, UrbanSound8K and the DCASE 2013 scene classification dataset (SCD) (e.g. evaluated on in \cite{cramer2019look}).
Others are: AudioSet \cite{gemmeke2017audio} which holds predominantly music and speech examples, and the newer DCASE datasets \cite{politis2020dataset}. These were not evaluated since we focused on comparison with a specific recent state-of-the-art method that does not use these datasets.

\textbf{Unsupervised representation learning.} Unsupervised approaches in the vision domain improve in great strides towards their supervised counterparts. MoCo \cite{he2020momentum}, SimCLR \cite{chen2020simple} and NPID \cite{wu2018improving} have shown to produce valuable features for downstream tasks. We base our contribution on NPID and extend it to the audio domain.

\textbf{Audio representation learning.} Compared to traditional hand-crafted features %
unsupervised training can lead to more robust and compact audio representations. 
\cite{lee2009unsupervised} applied deep belief networks to learn audio representations for speech and music. Generative methods have been explored in \cite{meyer2017unsupervised, chung2016audio, xu2017unsupervised} using variants of autoencoders. Audio representation learning has also been studied for speech \cite{hannun2014deep} and music \cite{thomee2016yfcc100m}. 

\textbf{Knowledge transfer from vision.} %
Transfer learning from visual tasks and exploiting audio-visual correspondences %
has been explored in the past. %
\cite{harwath2016unsupervised} was able to learn associations between free form audio, i.e. spoken sentences, and related images. By predicting simply if parts of an audio and video correspond, \cite{arandjelovic2017look} was able to learn good audio and visual representations. They were also able to extend this to localize sounding objects in a scene \cite{arandjelovic1712objects} and their approach was extended to be used in audio classification \cite{cramer2019look}.

Concurrent work which we use as basis for our evaluation improves downstream task performance by directly fine-tuning ImageNet pre-trained visual models. By mapping spectrograms to the format of a color image and by using a pre-trained Resnet50 they achieve state of the art results on audio event detection \cite{guzhov2020esresnet}.
However, these methods rely on a large network to leverage visual input for training. We present a method which can be used to pre-train networks optimized for the audio domain and low resources.

\vspace{-0.3cm} 
\section{Discriminative Learning of Sounds (DLS)}

Here we show how to pre-train ESResNet using DLS on four datasets and fine-tune on each dataset. We use the best performing ESResNet with attention for all experiments and pre-training steps.

\begin{figure}
    \centering
    \includegraphics[width=\columnwidth]{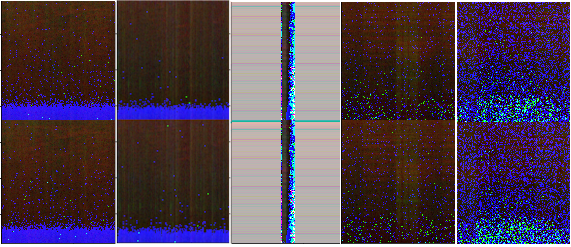}
    \caption{Spectrograms after pre-processing. Frequency ranges are split into three and concatenated as color channels. For illustration stereo channels are separated into rows. The leftmost two columns show \textit{children playing}, followed by \textit{car horns}. Large appearance changes exist within some classes.}
    \label{fig:spectrograms}
\end{figure}

\textbf{Audio Datasets.}
Most sounds in the datasets contain audio events recorded in natural environments, such as glass breaking or dog barking. DCASE2013 sounds also include longer recordings, such as riding in a bus. Across datasets, sounds differ in length and number of classes, summarized in table \ref{tab:datasets}.  Some contain events only on a fraction of their length, such as a single \textit{dog bark}, compared to other files filling the entire standardized length, such as \textit{kids playing} (see fig.  \ref{fig:spectrograms}).

\begin{table}[h]
    \centering
    \begin{tabular}{p{0.13\columnwidth}| p{0.12\columnwidth}|p{0.13\columnwidth}|p{0.13\columnwidth}|p{0.23\columnwidth}}
    \toprule
     & US8K    & ESC 50    & ESC 10 & DCASE 2013 \\
\toprule
\midrule
\makecell[l]{Events}   & 8732 & 2000 & 400 & 200\\
\midrule
\makecell[l]{Classes}        & 10   & 50 & 10 & 10\\
\midrule
\makecell[l]{Length}  &  $\leq$ 4s & 5s & 5s & 30s\\
\midrule
\makecell[l]{Fold}  &  1 of 10 & 1 of 5 & 1 of 5 & 1 of 2\\
\bottomrule
    \end{tabular}
    \caption{Dataset setup. Classification is hardest on ESC50 because it has the least data per class. We tune hyperparameters on one fold and use all data for unsupervised pre-training.}
    \label{tab:datasets}
\end{table}

\textbf{Spectrogram Network Input.} Power spectrograms, created with Short-Time Fourier Transform (STFT), are the input for all stages of training, following the method from \cite{guzhov2020esresnet}. We follow this method exactly to compare on equal grounds. Magnitude and phase are squared separately and the results are added. Spectrograms are divided into three equal sized parts to separate the higher, middle and lower frequencies. Finally the three parts are concatenated along a new channel dimension to create color images from the spectrograms to be processed by the network. 

%

\vspace{-0.3cm} 
\section{Network Pre-Training.} The pre-text task of instance discrimination generates features, useful for the downstream task of audio classification. The network's task is to assign a unique id to every spectrogram. Applied on images, this creates a feature space where \textit{visually} similar images are grouped. However, visual similarity is not an explicitly defined criteria in the loss function, but a by-product of the pressure on the network to structure and differentiate between all images of the dataset. 


We follow \cite{wu2018unsupervised} and use spectrograms as input. We train ESResNet $f_\theta$ with weights $\theta$ to map a spectrogram $x_i$ with the assigned id $i$ in the training set to a latent output vector $\mathbf{v_i}$. The vector is L2 normalized, so $\vert\vert\mathbf{v_i}\vert\vert = 1$ and updates its entry in a memory bank in each training iteration to calculate $\{\mathbf{v}_j\}$ of all spectrograms. This allows efficient calculation of a non-parametric softmax to get the assigned instance id probabilities.
The probability of a spectrogram $x$ mapped to vector $\mathbf{v}$ having training set id $i$ is defined by:
\begin{equation}
    P(i|\mathbf{v}) = \frac{\exp{(\mathbf{v}_i^T \mathbf{v}/\tau)}}{\sum^n_{j=1}\exp{(\mathbf{v}_j^T \mathbf{v}/\tau)}}
    \label{eqn:softmax}
\end{equation}
where $\tau$ is the Softmax temparature parameter. Stochastic gradient decent is used to minimize the log-likelihood: $ \argmin_{\theta} -\sum_{i=1}^n \log{P(i|f_{\theta}(x_i))}$ during training.

The final feature mapping has the advantage that similar spectrograms are mapped close in the feature space. This enables to group sounds with similar characteristics, as sketched in fig. \ref{fig:schema_training}. While this enables our performance on the downstream task, larger variance within a class than between classes is an issue and limits this approach. Examples of very different car horns can be seen in fig. \ref{fig:spectrograms}.

\textbf{Evaluation of Unsupervised Learning.} We first trained and evaluated on each dataset separately, according to the official training/evaluation folds (see tab. \ref{tab:datasets}). Every instance of the evaluation set is mapped and we check its label alignment with its weighted K=5 nearest neighbors in the training data feature space. We tune hyperparameters separately for each dataset but found the optimal configuration was almost the same for all. Therefore, we use the average parameters: \textit{Embedding dimension}=128, \textit{NCE-K}=64 and \textit{NCE-T}=0.4, which are the number of negative samples and Softmax temperature of the Noise Contrastive Estimation in \cite{wu2018unsupervised}. We trained 200 epochs with a batchsize of 64.


In a second phase, we use all available audio datasets and folds to combine one large training dataset, without evaluation or test set. Therefore we train the embedding with the fixed hyperparameters found in phase one and no longer evaluate the performance in this step. However, we observe that the training loss at epoch 200 is sufficiently converged. The result of this phase is the pre-trained ESResNet.
\section{Network Fine-Tuning.}
\begin{figure*}
\begin{center}
\begin{tabular}{cc}
    \includegraphics[width=0.9\columnwidth]{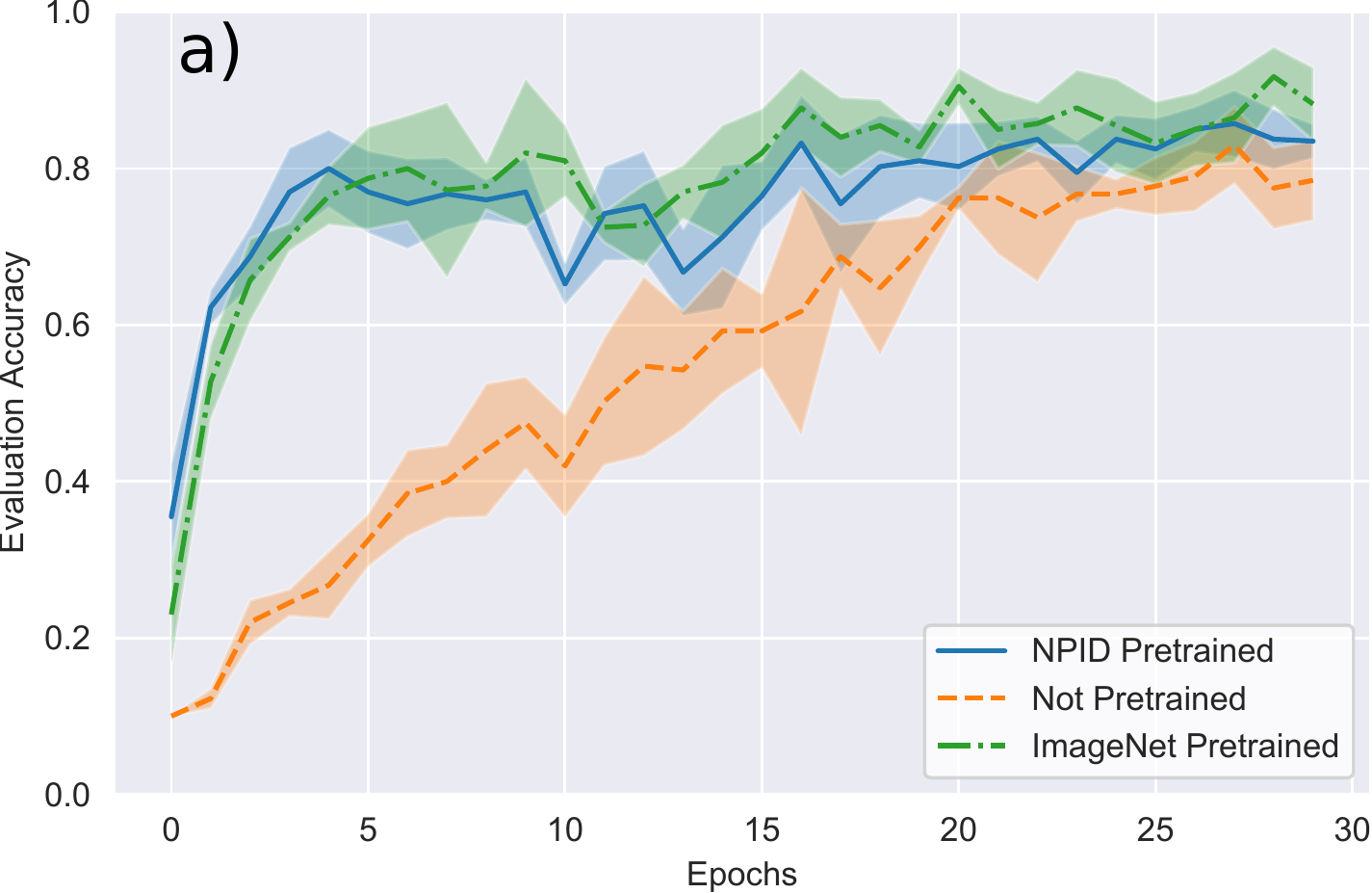} & \includegraphics[width=0.9\columnwidth]{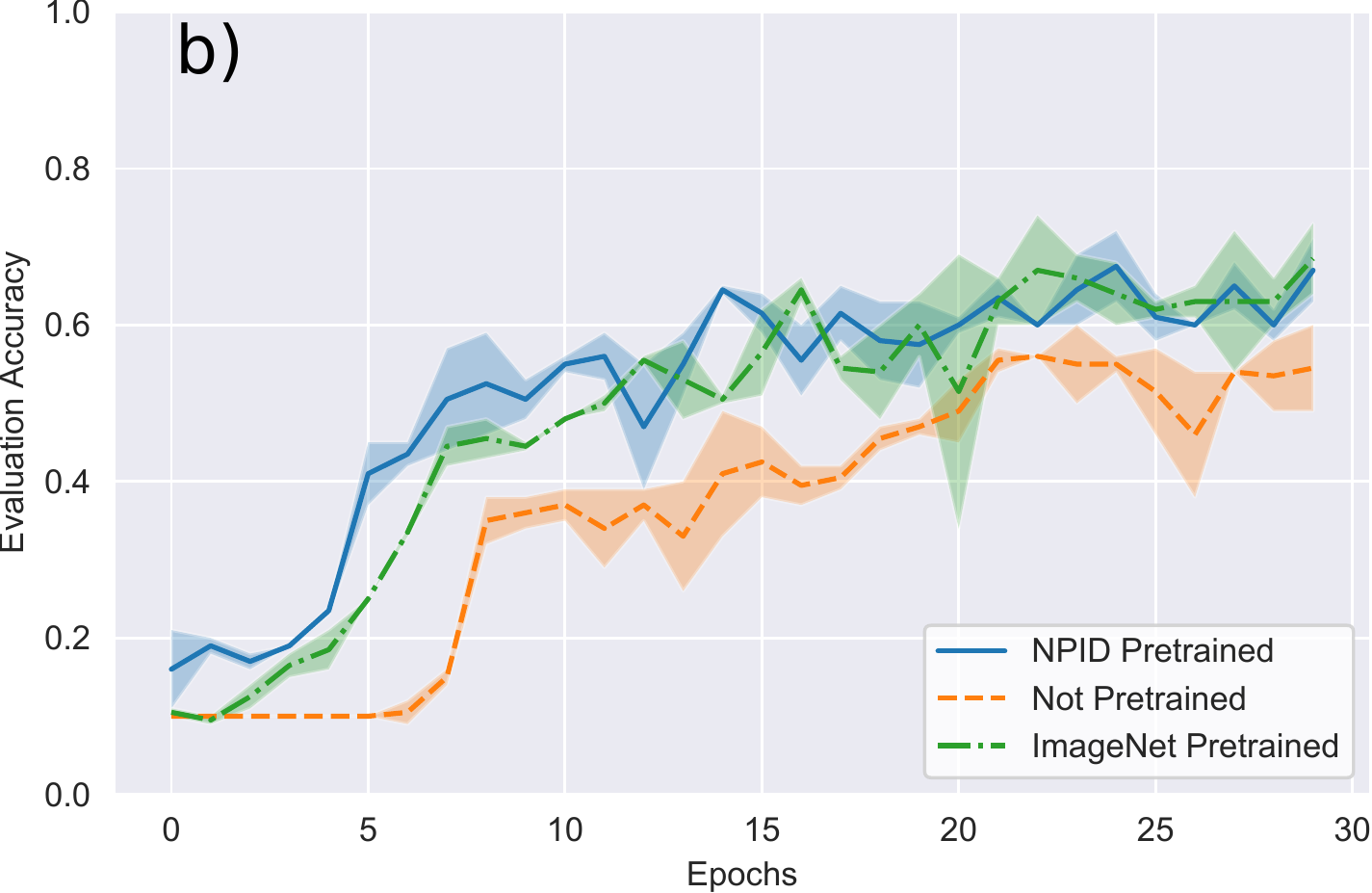} \\
    \includegraphics[width=0.9\columnwidth]{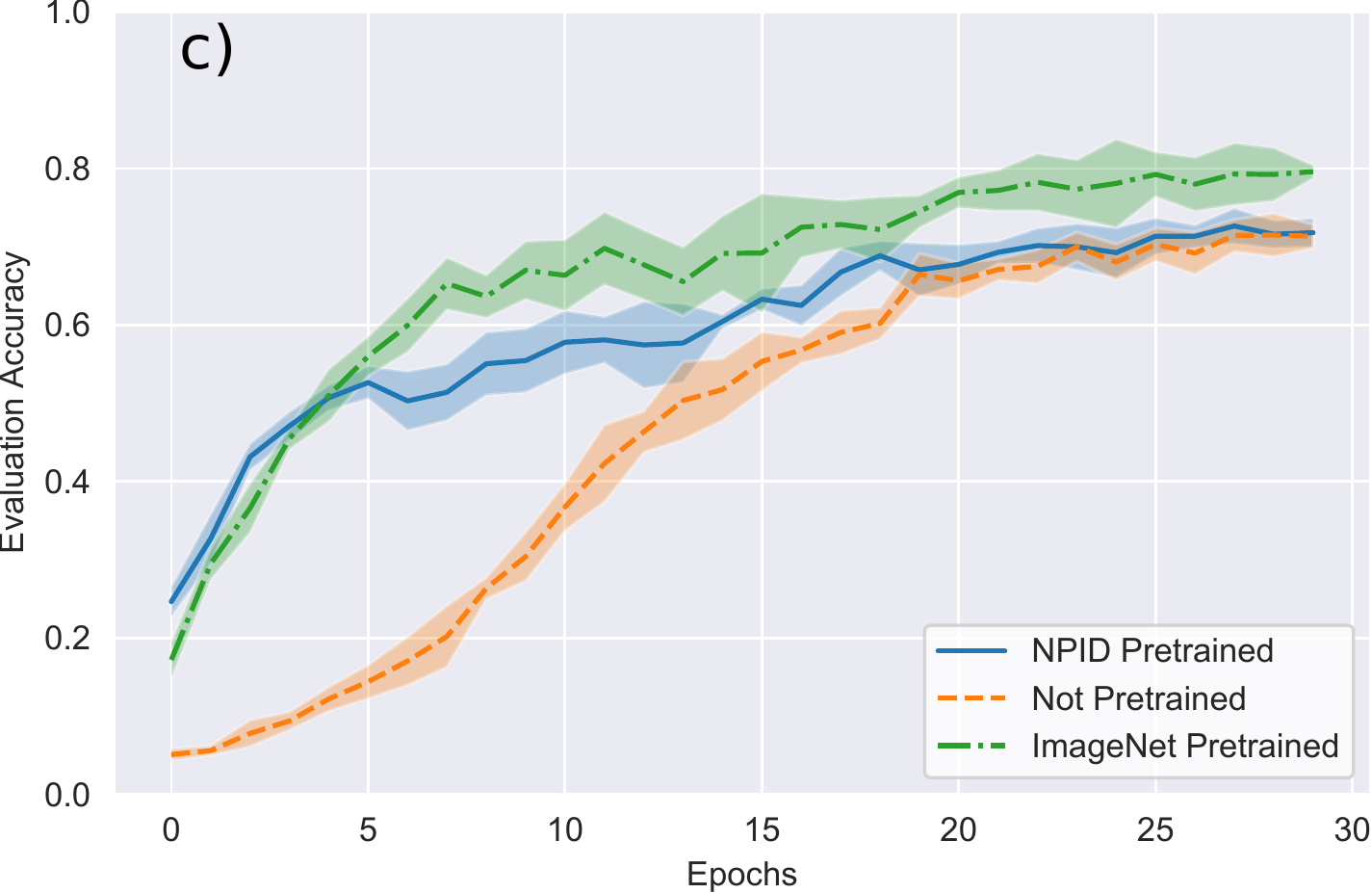} & 
    \includegraphics[width=0.9\columnwidth]{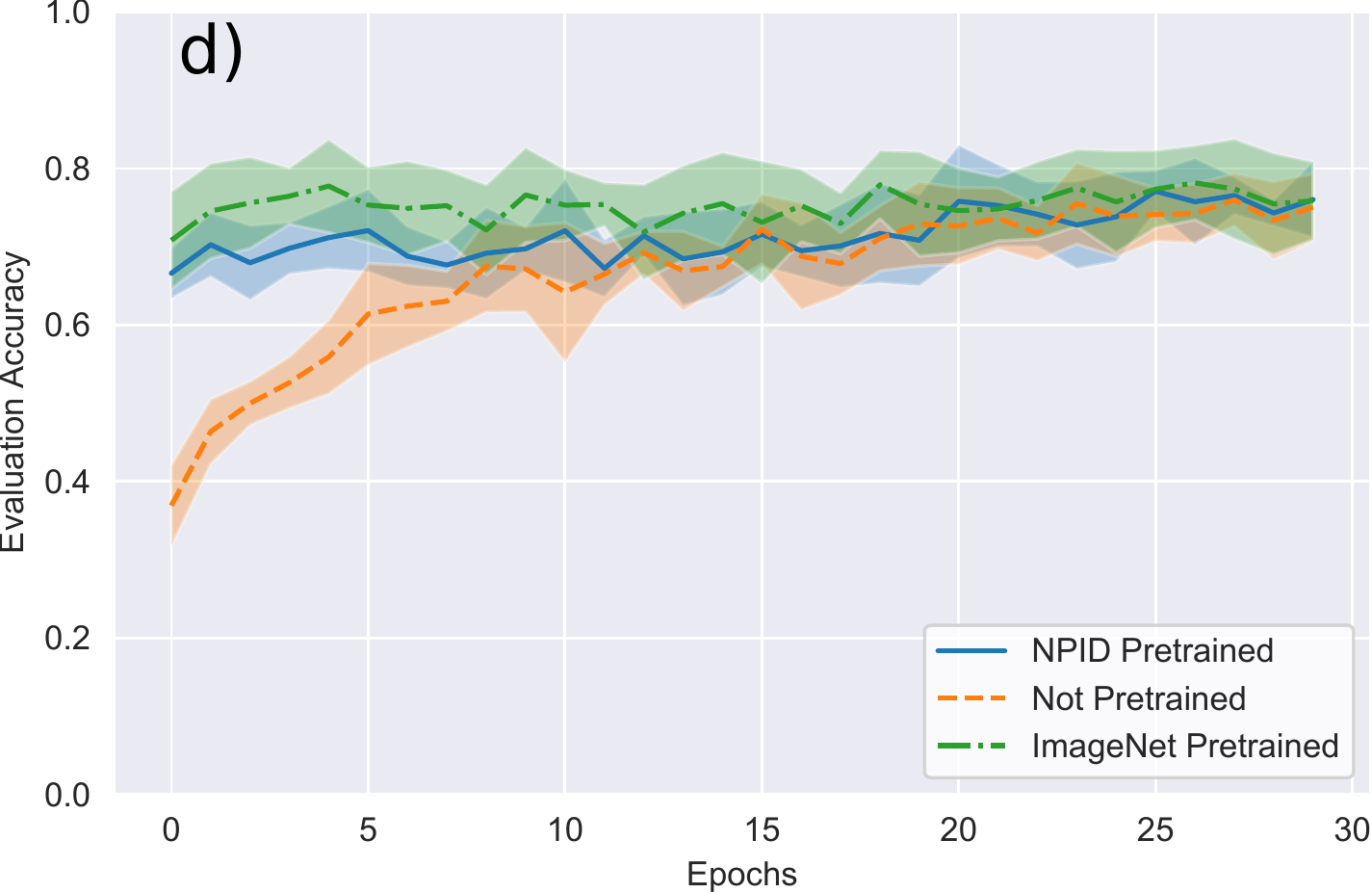} \\
\end{tabular}
\end{center}
\caption{Evaluation accuracy over epochs on a) ESC10, b) DCASE2013, c) ESC50 and d) UrbanSounds8k. Shaded areas show standard deviation of average n-fold cross validation results. The average accuracy pretrained with DLS is on par with ImageNet pre-training and significantly better than training from scratch. Even though we use only sound data with a fraction of the size of ImageNet, we reach a similar performance gain.
In d) pre-training shows on par results with ImageNet pretraining. In c) DLS pre-training outperforms from scratch training over the first 18 epochs.
In b) DLS pre-training even outperforms ImageNet pre-training on the first 11 epochs.
}
\label{fig:plots}
\begin{multicols}{2}

\vspace{-0.3cm} 
\section{Experimental Results}
We evaluated the performance of ESResNet with attention with DLS pretraining, Imagenet pretraining and training from scratch. The architectures are fully identical \cite{guzhov2020esresnet}.

\textbf{DCASE2013 and ESC10.} Training on small datasets benefits largely from pre-training, as can be seen by the performance gain on  ESC10 and DCASE2013 (fig. \ref{fig:plots}). On ESC10, DLS pre-training outperforms ImageNet pretraining slightly in the first three epochs, rising to 80\% accuracy where training from scratch remains under 40\%. On DCASE2013, DLS pre-training is best for 11 epochs, rising towards 50\% accuracy and continuing on par with ImageNet pre-training. 

\textbf{ESC50 and UrbanSounds8k.} On larger datasets, the accuracy gain in the first epochs is similar between DLS and ImageNet pre-training. However DLS was trained in significantly shorter time and on less data. On ESC50, it outperforms from scratch training significantly and rises to over 60\% accuracy in the first 15 epochs. UrbanSounds8k is the largest dataset and both pre-training methods show similar performance gain per epoch. Each epoch contains more sound files than in any other dataset so methods are already close after the first epoch when they are evaluated. DLS still outperforms training from scratch on 5-6 epochs.

\textbf{Densenet as backbone.} We also compared using Densenet as backbone and omit graphs for brevity. Results show slower performance gain over the first 15 epochs and then level off at similar levels. All trends are similar though on DCASE2013, all performance gains are very close before epoch 30.

\vspace{-0.3cm} 
\section{Discussion}
We presented an unsupervised pre-training method which enables fast fine-tuning when training networks for audio event classification. While using only a comparatively small amount of sound data, we show that we can gain early performance as the same network pre-trained on ImageNet. We validate our results on three commonly used datasets for audio classification. Our approach enables network training on devices with limited computing resources, e.g. for continuous improvement by adopting novel collected data. Because pre-training with DLS is unsupervised, novel unlabeled data can be integrated with existing data seamlessly.

\end{multicols}

\end{figure*}

The pretrained ESResNet is fine-tuned and evaluated on each dataset individually. %
The final classification layer is set to the number of classes depending on the dataset we fine-tune on, which means it needs to be retrained. We found fine-tuning all layers yields faster performance gain than fixing them and only fine-tuning the classifier. Results are shown in fig. \ref{fig:plots}.

\textbf{Comparison with ImageNet Data}
We compare pre-training by image classification on ImageNet with DLS on the four audio datasets: DCASE2013 (SCD), ESC-50/ESC-10 and UrbanSounds 8k which provide 9.3Gb of sound data. ImageNet consists of over 200Gb of data, with 14,197,122 images.
Training a classification task on ImageNet with Resnet50, which is also the backbone of the ESResNet architecture, can take almost two weeks on a single GPU \cite{you2018imagenet}. 
Even if we assume ImageNet training happens on a similar modern setup as ours (4x GeForce GTX 1080 Ti), the training time of ImageNet would still be at least 1-2 days. Training DLS with ESResNet on 200 epochs takes 4 hours.  

\clearpage


\bibliographystyle{IEEEbib}
\bibliography{our_bib}

\end{document}